\documentclass[12pt,a4paper]{iopart}
\pdfoutput=1
\usepackage{iopams}  

\usepackage{graphicx}
\usepackage{color}
\usepackage{cite}

\expandafter\let\csname equation*\endcsname\relax
\expandafter\let\csname endequation*\endcsname\relax
\usepackage{amsmath}

\usepackage{dsfont}

\begin{document}

\title{Characterization of Anisotropic Gaussian Random Fields by Minkowski Tensors}

\author{Michael Andreas Klatt$^{1,2}$\footnote{Present address:
Institut f\"ur Theoretische Physik II: Weiche Materie, 
Heinrich-Heine-Universit\"at D\"usseldorf 40225 D\"usseldorf, Germany}, 
Max H\"ormann$^{1}$ and Klaus Mecke$^1$}

\address{$^1$ Friedrich-Alexander-Universit\"at Erlangen-N\"urnberg (FAU),
  Institut f\"ur Theoretische Physik, Staudtstr. 7, 91058 Erlangen,
  Germany; $^2$ Experimental Physics, Saarland University, Center for 
  Biophysics, 66123 Saarbrücken, Germany}

\ead{michael.klatt@fau.de}
\vspace{10pt}
\begin{indented}
\item[]\today
\end{indented}

\newcommand\N{\mathbb{N}}
\newcommand\R{\mathbb{R}}
\newcommand\E{\mathbb{E}}
\renewcommand\P{\mathbb{P}}
\newcommand\Var{\mathbb{V}\mathrm{ar}}
\newcommand\Cov{\mathbb{C}\mathrm{ov}}
\newcommand\psd{s}

\newcommand\exponent[1]{e^{#1}}

\newcommand\f{f}
\newcommand\g{g}
\newcommand\F{F}
\newcommand\G{G}

\newcommand\ind[1]{\ifiopams\mathds{1} \fi\{#1\}}

\newcommand\Ob{\Omega}
\newcommand\egi{\psi}
\newcommand\egiK{\Psi_K}
\newcommand\egiC{\psi}

\begin{abstract}
Gaussian random fields are among the most important models of amorphous 
spatial structures and appear across length scales in a variety of 
physical, biological, and geological applications, from composite 
materials to geospatial data.
Anisotropy in such systems can sensitively and comprehensively be 
characterized by the so-called Minkowski tensors from integral geometry.

Here, we analytically calculate the expected Minkowski tensors of 
arbitrary rank for the level sets of Gaussian random fields.
The explicit expressions for interfacial Minkowski tensors are confirmed 
in detailed simulations.
We demonstrate how the Minkowski tensors detect and characterize 
the anisotropy of the level sets, and we clarify which shape information 
is contained in the Minkowski tensors of different rank.

Using an irreducible representation of the Minkowski tensors in 
the Euclidean plane, we show that higher-rank tensors indeed contain 
additional anisotropy information compared to a rank two tensor.
Surprisingly, we can nevertheless predict this information from the 
second-rank tensor if we assume that the random field is Gaussian.
This relation between tensors of different rank is independent of the 
details of the model.
It is, therefore, useful for a null hypothesis test 
that detects non-Gaussianities in anisotropic random fields.
\end{abstract}

\vspace{2pc}

\maketitle

\section{Shape analysis of anisotropic random fields}

\begin{figure}[t]
  \centering
  \includegraphics[width=0.61\linewidth]{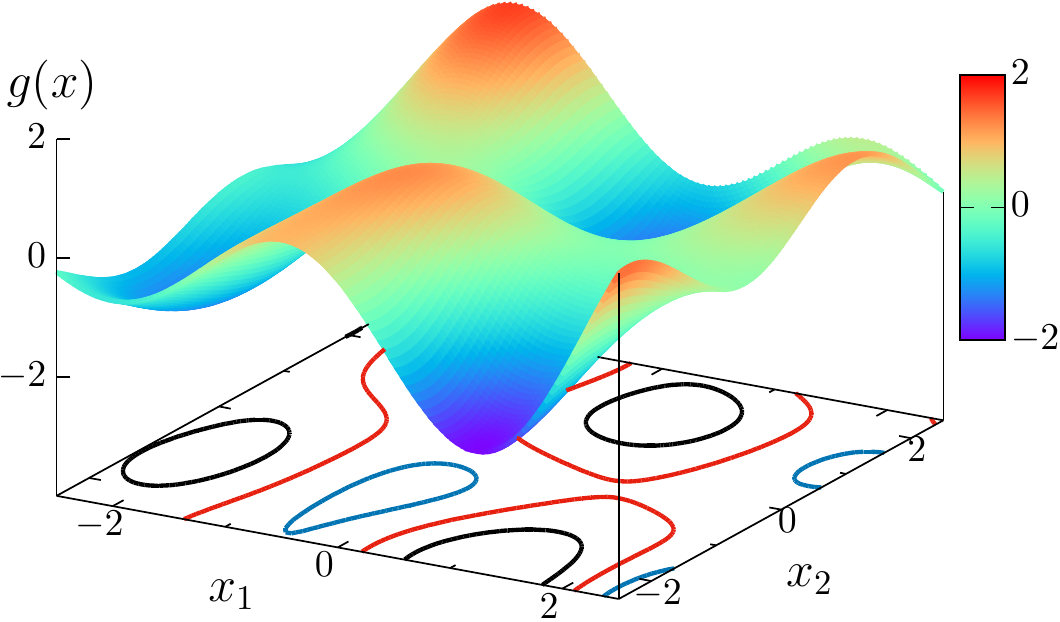}
  \includegraphics[width=0.37\linewidth]{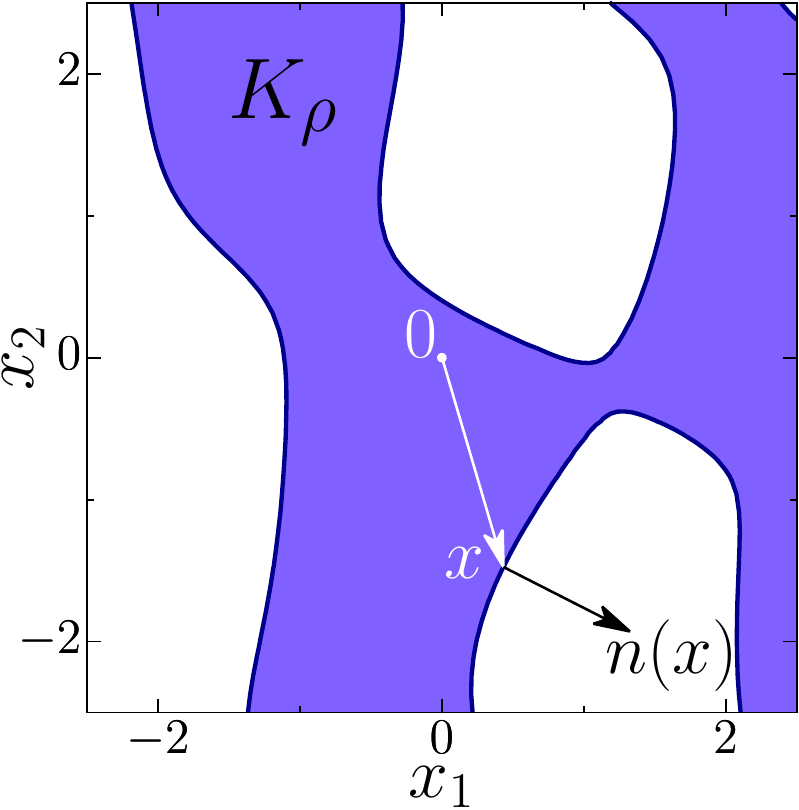}
  \caption{An exemplary realization of an anisotropic Gaussian random 
  field $\g(x)$ for $x=(x_1,x_2)\in\R^2$ is shown on the left hand side 
  together with level sets for thresholds $\rho=-1,0,1$. On the 
  right-hand side, the corresponding excursion set $K_{\rho}$ is shown 
  for $\rho=0$. A normal vector $n(x)$ at a position $x$ on the boundary 
  is also depicted.} 
  \label{fig_grf_sample}
\end{figure}

Intuitively speaking, a random field is a function with random 
functional values, see Fig.~\ref{fig_grf_sample}.
More precisely, the random field assigns to each point in space a random 
variable, typically, such that the functional values of proximate points 
are correlated with each other~\cite{adler_random_2007, 
chiu_stochastic_2013, vanmarcke_random_2010}.
Hence, random fields are versatile models that are essential to a broad 
range of fields in physics, including quantum 
mechanics~\cite{berry_regular_1977}, 
cosmology~\cite{peebles_principles_1993}, and material 
sciences~\cite{torquato_random_2002}.
They can be used to model random intensities or temperature 
profiles~\cite{liddle_cosmological_2000, Goodman2007speckle, 
dennis_nodal_2007, klatt_detecting_2019, collischon_tracking_2021},  
coarse-grained particle density profiles~\cite{scholz_direct_2015, 
schuetrumpf_appearance_2015, mizuno_spatial_2016, zheng_hidden_2021, 
jiao_hyperuniformity_2021},  rough 
surfaces~\cite{whitehouse_surface_2000, ji_ultra-low-loss_2017, 
spengler_strength_2019},  or electric field 
distributions~\cite{BerryDennis2000, edagawa_photonic_2008, 
klatt_gap_2021}, to name just a few examples.

The arguably most important class of random fields are Gaussian random 
fields~\cite{adler_geometry_1981, vanmarcke_random_2010}.
Their importance is essentially due to the central limit theorem 
because a sum of many independent and identically distributed 
fluctuations can, virtually always, be well approximated by a Gaussian 
random field.
Moreover, Gaussian random fields are specified by only their first and 
second moments, which leads to simple and controllable models in 
applications.
Gaussian random fields have been used to model, for example, chaotic 
quantum systems~\cite{berry_regular_1977, berry_statistics_2002}, random 
speckle patterns in optics~\cite{Goodman1985statistical} 
microemulsions~\cite{teubner_level_1991}, rough 
surfaces~\cite{mantz_utilising_2008, filliger_3d_2012, 
lessel_impact_2013}, thin films~\cite{becker_complex_2003}, 
heterogeneous materials~\cite{torquato_random_2002}, sea 
waves~\cite{LonguetHiggins1957}, and prominently the density 
fluctuations in the early universe~\cite{peebles_principles_1993}.

An efficient way to analyze the complex shape of realizations of 
(Gaussian) random fields and capture higher-point information beyond 
the two-point correlation function is to characterize the corresponding 
level sets.
More precisely, the random field is turned into a two-phase medium via 
thresholding.
The set of all points with a functional value below the threshold is 
called the \textit{excursion set}~\cite{adler_geometry_1981}.
The geometry of the excursion set is then typically studied as a 
function of the threshold.
Notably, the ``density profile'' of a random field can thus 
also be used to model porous media~\cite{teubner_level_1991, 
roberts_transport_1995, RobertsTorquato1999}.

Minkowski functionals from integral geometry (also known as intrinsic 
volumes) are a powerful set of shape descriptors of excursion 
sets~\cite{schroder-turk_minkowski_2011, chiu_stochastic_2013, 
adler_random_2007}.
They comprise all `additive shape information' according to the Hadwiger 
theorem~\cite{schneider_stochastic_2008}.
As robust and versatile structural characteristics of random spatial and 
planar structures, they have been successfully applied to a broad range 
of applications~\cite{schroder-turk_minkowski_2011, 
klatt_morphometry_2016}, from 
astronomy~\cite{kerscher_morphological_2001, goring_morphometric_2013, 
schuetrumpf_time-dependent_2013, ebner_goodness--fit_2018, 
klatt_detecting_2019, collischon_tracking_2021}
to rough surfaces~\cite{mecke_robust_1994, 
becker_complex_2003,spengler_strength_2019}, biological 
tissues~\cite{barbosa_integral-geometry_2014, rath_strength_2008}, or 
porous media~\cite{arns_3d_2010, scholz_direct_2015, 
klatt_anisotropy_2017}.

There is a rich literature on the characterization of Gaussian excursion 
sets~\cite{adler_geometry_1981, vanmarcke_random_2010, BerryDennis2000, 
dennis_nodal_2007, torquato_hyperuniformity_2016, taylor_vortex_2016}.
Specifically, the Minkowski functionals (MFs) for isotropic Gaussian 
random fields have already been intensively studied, including mean 
values~\cite{tomita_statistical_1986, mantz_utilising_2008, 
adler_random_2007}, second moments, and asymptotic 
distributions~\cite{estrade_central_2016, 
marinucci_non-universality_2016, cammarota_quantitative_2016, 
nourdin_nodal_2017, paper:Mueller, paper:KratzVadlamani, 
di_bernardino_test_2017}.
The MFs have also been extensively used to study whether the cosmic 
microwave background is a Gaussian random field, e.g., 
see~\cite{schmalzing_disentangling_1999, raeth, 
ducout_non-gaussianity_2013, novaes_local_2016, 
planck_collaboration_planck_2020}.

There are considerably fewer studies of the anisotropy of Gaussian 
random fields.
Examples include nonisotropic quantum systems~\cite{UrbinaRichter2003}, 
anisotropic heterogeneous materials~\cite{torquato_random_2002, 
BonamiEstrade2003}, friction and rough surfaces~\cite{filliger_3d_2012}, 
the modeling of sea waves~\cite{LonguetHiggins1957, 
estrade_anisotropic_2020}, and the stochastic heat 
equation~\cite{Xiao2009}.
Fractal anisotropic Gaussian random fields and their shape have been 
analyzed~\cite{BonamiEstrade2003, cheng_mean_2016}
for models of porous materials~\cite{BonamiEstrade2003} and biological 
tissues~\cite{richard_statistical_2010}.
The mean values of MFs were also derived for anisotropic Gaussian random 
fields~\cite{adler_random_2007}, and MFs were applied to analyze 
anisotropic porous materials~\cite{arns_3d_2010} and 
surfaces~\cite{filliger_3d_2012}.

The scalar MFs, however, do not directly detect or measure anisotropy, 
neither the degree nor the preferred direction.
For this purpose, the MFs have been extended to the so-called Minkowski 
tensors (MTs)~\cite{McMullen1997}.
The MTs extend the notion of volume, surface area, and curvature to 
tensorial morphometric measures~\cite{schneider_stochastic_2008, 
schroder-turk_minkowski_2011, klatt_cell_2017}.
Thus, the MTs allow for a sensitive and comprehensive anisotropy analysis with 
respect to different geometrical properties, like surface area, 
circumference, or curvature~\cite{schneider_stochastic_2008, 
schroder-turk_minkowski_2011, bobel_kinetics_2016, klatt_mean-intercept_2017-1, 
klatt_mean-intercept_2017}.
Because they are robust against noise and their computation time grows 
linearly with the system size, they are efficient and comprehensive 
shape descriptors for data analysis of random spatial structures in 
experiments~\cite{schroder-turk_minkowski_2011}.
Minkowski tensors have already been successfully applied to a variety of 
experimental data sets, including solid 
foams~\cite{schroder-turk_minkowski_2011, saadatfar_structure_2012} and 
granular packings~\cite{schroder-turk_minkowski_2013, 
schaller_non-universal_2015, xia_structural_2015, WeisSchroeter2017}.
Recently, MTs were applied to analyze isotropic Gaussian random fields 
on the sphere and the cosmic microwave 
background~\cite{chingangbam_tensor_2017, kapahtia_novel_2017, 
ganesan_tensor_2017}, as well as two-dimensional slices of the dark 
matter density field~\cite{appleby_minkowski_2017}, see 
also~\cite{appleby_minkowski_2018, chingangbam_geometrical_2021}.

Here, we use the MTs to characterize the anisotropy of Gaussian random 
fields~\cite{Hoermann2014, klatt_morphometry_2016}; see 
Sec.~\ref{sec:RFMT} for a brief introduction to random fields and 
Minkowski tensors.
We show how the MTs comprehensively detect and quantify anisotropy in 
general random fields, see Sec.~\ref{sec:general}.
A key result is the mean-value formula for interfacial anisotropy of 
general smooth random fields in Eq.~\eqref{eq:intMT_B}.
We explicitly evaluate this formula for surface tensors of arbitrary rank 
and Gaussian random fields in arbitrary dimensions in 
Sec.~\ref{sec:GRF}, see Eq.~\eqref{eq:w10s-in-d-dim}.

We pay special attention to the question which information is contained 
in MT of different order and rank.
Surprisingly, we find that while higher-rank MTs indeed contain 
additional information, there is a strict relation between MT of 
different rank for a Gaussian random field independent of its 
correlation function.
In fact, only assuming that the random field is Gaussian, MTs of 
arbitrary rank can be robustly predicted from the second-rank MT.
A comparison of these predicted and measured values allows for a null 
hypothesis test that detects non-Gaussianities in anisotropic random 
fields.
An outlook is given in Sec.~\ref{sec:conclusion}.

We demonstrate the robustness of our MT analysis to finite-size effects 
and statistical noise by detailed simulations that accurately agree with 
our analytic predictions. 
Therefore, we define a class of parametric models with a tunable degree 
of anisotropy in \ref{sec:RW}.
Finally, we provide an introduction to \textit{irreducible Minkowski 
tensors} (IMT) in \ref{sec:irr} that is used in our analysis of Gaussian 
random fields, but that may be useful for more general planar random 
structures.


\section{Background on random fields and Minkowski tensors}
\label{sec:RFMT}

We here restrict our analysis to real-valued random fields that are 
defined on the Euclidean space $\R^d$.
More precisely, such a random field can be defined as a collection of 
random variables so that it assigns to each site $x\in\R^d$ a 
real-valued random variable~\cite{adler_random_2007}.
The functional value is sometimes called the `state' of the system and 
$x$ is a `site', `index', or `parameter'~\cite{vanmarcke_random_2010}.
In the following, we denote a realization of a general random field by a 
function $\f(x)$ and that of a Gaussian random field by a function 
$\g(x)$.
We denote the corresponding random variables using capital letters, 
i.e., $\F(x)$ and $\G(x)$.

\subsection{General random fields}

The distribution of a random field $F$ is specified by its 
\textit{finite-dimensional distributions}, i.e., the joint distributions 
of $\F(x_1)$, \ldots, $\F(x_n)$ for all $x_1$, \ldots, $x_n\in\R^d$ and 
all $n\in\N$.
In the following, we assume that the random field is 
\textit{statistically homogeneous} (or \textit{stationary}), i.e., we 
assume that the finite-dimensional distributions are invariant under 
translations of the index set.
Intuitively speaking, this implies that there is no preferred choice for 
the origin. A translation of the system does not change the statistical 
properties of the random field.
Moreover, we assume that the fields are square-integrable (so that the 
first and second moments are well defined), ergodic, and sufficiently 
differentiable (so that all integral expressions are well defined, and 
limit operations can be exchanged).

Statistical homogeneity implies that the \textit{mean function} of the 
random field $\mu(x) := \E[\F(x)]$ is a constant for all $x\in\R^d$.
Without loss of generality, we here choose $\mu(x) \equiv 0$.
The \textit{covariance function} of the random field $F$ is then defined by:
\begin{eqnarray}
  \Cov(x,y) := \E[\F(x)\F(y)]
\end{eqnarray}
for $x,y\in\R^d$.
Because of the statistical homogeneity, $\Cov(x,y)$ only depends on the difference of $x$ and $y$, 
so that we can use the shorthand notation $\Cov(x-y):=\Cov(x,y)$.
Since we are free to rescale the functional values, we here choose, 
again without loss of generality, $\Var[\F(0)]=\Cov(0)=1$.
The Fourier transformation of the covariance function is the 
\textit{power spectral density}:
\begin{eqnarray}
  \psd(q) = \int_{\R^d}\Cov(x)\exponent{-i\langle{}q,x\rangle}dx,
\end{eqnarray}
where $q\in\R^d$ is called the wave vector, and $\langle{}q,x\rangle$ denotes the scalar product.

Finally, we define the \textit{excursion set} $K_{\rho}$
for a given realization $\f(x)$ of the random field $\F$:
\begin{eqnarray}
  K_{\rho} &:=& \{x\in\R^d\mid \f(x) \geq \rho\},
  \label{eq:def_excursion_set}
\end{eqnarray}
i.e., $K_{\rho}$ is the set of all points $x$ for which the functional 
value $\f(x)$ is larger or equal to a given threshold 
$\rho$~\cite{adler_geometry_1981}.
Its boundary $\partial K_{\rho}=f^{-1}(\rho)$ is known as a \textit{level 
set}, see Fig.~\ref{fig_grf_sample}.

\subsection{Gaussian random fields}
\label{sec:grf}

If all finite-dimensional distributions are Gaussian, then the random 
field $\G$ is said to be a \textit{Gaussian random field}.
Recall that the distribution of a Gaussian random vector $Y\in R^n$ is 
determined by its mean vector $\mu_Y:=\E[Y]$ and covariance matrix 
$C_Y:=\Cov[Y]$.
If $C_Y$ is regular, then $Y$ has the probability density %
\begin{equation}
  \varphi_Y(y):=\frac{1}{(2\pi)^{n/2}\sqrt{\det 
  C_Y}}e^{-\frac{1}{2}\langle y-\mu_Y,C_Y^{-1}(y-\mu_Y)\rangle}.
\end{equation}
Similar to a multivariate Gaussian distribution, the distribution of a 
stationary Gaussian random field is determined by $\mu$ and $\Cov(x)$.
For our choice of units, the probability density function of $G(0)$ is a 
standard normal distribution, i.e., centered and with unit variance.

For a smooth, centered Gaussian random field, the gradient $\nabla G(0)$ 
also has a centered Gaussian distribution~\cite[Corollary 
4.4]{potthoff_sample_2010}, see also \cite[Lemma 
2.3.6]{sasvari_multivariate_2013}.
For a statistically homogeneous, differentiable random field, the 
functional value and the gradient are 
uncorrelated~\cite[Eq.~(5.5.6)]{adler_random_2007}.
Hence they are even stochastically independent for a Gaussian random 
field $\G$.

In the following, we study functionals $M:\g\rightarrow\R$ that map the 
realization of a random field to a real number, like the volume of the 
intersection of an excursion set with an observation window.
We denote the mean value of such a functional by $\mathbb{E}[M]$.

\subsection{Minkowski tensors}

Minkowski functionals (MF) and Minkowski tensors (MT) have a rigorous, 
mathematical foundation in integral 
geometry~\cite{schneider_stochastic_2008, chiu_stochastic_2013, 
adler_random_2007}, which also offers insights for applications.
For example, Hadwiger's theorem~\cite{hadwiger_vorlesungen_1957} 
guarantees that the MFs contain all additive scalar shape information 
(that is, any motion invariant functional that is additive and
fulfills minor continuity requirements is a linear combination of MFs).
This remarkable statement has been generalized to the 
MTs~\cite{Alesker:1999a,alesker_description_1999}.
Moreover, the MFs and MTs have intuitive geometric interpretations, 
including volume, surface area, and moments of the distributions of mass 
or normal vectors~\cite{schroder-turk_minkowski_2011, 
klatt_mean-intercept_2017}.

For a smooth, bounded domain $K\subset\R^d$, the MFs can be represented 
by integrals over the volume or as curvature-weighted integrals over the 
boundary $\partial K$, respectively:
\begin{eqnarray}
  W_0 &:=& \int_K dx,\\
  W_{\nu} &:=& \int_{\partial K} H_{\nu-1}(K,x) dS(x), \quad \mathrm{for\ } \nu={1,\ldots,d},
  \label{eq_def_MF}
\end{eqnarray}
where $dS$ denotes the $(d-1)$-dimensional surface measure, and the 
$H_j(K,x)$ are the elementary symmetric polynomials of the 
\textit{principal curvatures} $\kappa_1(K,x), \ldots, \kappa_{d-1}(K,x)$ 
of $K$ at $x\in\partial K$:
\begin{equation}
  H_j(K,x) := \sum_{|I|=j}\prod_{l\in I}\kappa_l(K,x)
\end{equation}
with the summation extending over all subsets $I\subset\{1,\ldots,d-1\}$ 
of cardinality $j$ and $H_0(K,x) = 1$.
Different normalizing constants are used in different context. Here, we 
choose a normalization such that $W_1 = \int_{\partial K} dS(x)$
is the surface area (or circumference) of $K$.

The scalar MFs can naturally be generalized to tensors that quantify 
anisotropy, simply by multiplying the integrand with the symmetric 
tensor product $x^rn^s(x)$ of the position vector $x$ and the normal vector 
$n(x)$ with $r,s\in\N$:
\begin{eqnarray}
  W_0^{r,0}(K)     &:=& \int_K x^r dx,\\
  W_{\nu}^{r,s}(K) &:=& \int_{\partial K} x^rn^s(x) H_{\nu-1}(K,x) dS(x), \quad \mathrm{for\ } \nu={1,\ldots,d},
  \label{eq_def_MT}
\end{eqnarray}
where the normal vector $n(x)$ is perpendicular to the (smooth) 
interface at position $x\in\partial K_\rho$ and pointing outwards, see 
Fig.~\ref{fig_grf_sample}.
For a mathematical definition of the symmetric tensor product, e.g., 
see~\cite{hug_schneider_2017}.
Here it suffices to focus on the tensors $W_0^{r,0}$ and $W_{\nu}^{0,s}$ 
and hence on the tensor product $y^t$ of a single vector $y\in\R^d$ and 
$t\in\N$, for which the Cartesian representation is simply given by a 
matrix with entries $(y^t)_{j_1\ldots j_s} = y_{j_1}\ldots y_{j_s}$.

The tensors $W_0^{r,0}$  quantify the distribution of mass in a 
homogeneous solid body.
The tensors $W_{\nu}^{0,s}$ describe how the orientation of the 
interface is distributed, possibly weighted by the curvature of the 
interface.
In contrast to $W_0^{r,0}$, the tensors $W_{\nu}^{0,s}$ are translation 
invariant.

\section{Anisotropy quantification for general random fields} 
\label{sec:general}

For a statistically homogeneous random field, we can study the Minkowski 
functionals and tensors of the excursion set $K_{\rho}$ in the 
thermodynamic limit.
More precisely, we first compute the mean value $\E[W_{\nu}^{r,s}(K_{\rho}\cap 
\Ob)]$ of a Minkowski tensor in a finite observation window 
$\Omega\subset\R^d$, then rescale the expectation with the size of the 
window, and finally consider the thermodynamic limit:
\begin{eqnarray}
  w_{\nu}^{r,s}(\rho) &:=& \lim_{|\Ob|\to\infty} 
  \frac{\E[W_{\nu}^{r,s}(K_{\rho}\cap \Ob)]}{|\Ob|^{1+r/d}}, 
\end{eqnarray} where $|\Ob|$ is a shorthand notation for the volume 
$W_0(\Ob)$ of the observation window.
The limit is a \textit{Minkowski tensor density}.
For example, the density of the volume is the well-known volume fraction 
$\phi(\rho)$ of the excursion set $K_{\rho}$:
\begin{eqnarray}
  \phi(\rho) := w_0^{0,0}(\rho)
  &=& \lim_{|\Ob|\to\infty} \frac{1}{|\Ob|} \E\left[\int_{\Ob}\ind{\F(x)\geq\rho}dx\right]\\
  &=& \lim_{|\Ob|\to\infty} \int_{\Ob} \frac{1}{|\Ob|} \E\left[\ind{\F(x)\geq\rho}\right] dx\\
  &=& \E\left[\ind{\F(0)\geq\rho}\right] \lim_{|\Ob|\to\infty} \int_{\Ob} \frac{1}{|\Ob|} dx\\
  &=& \P\left[\F(0)\geq\rho\right]
\end{eqnarray}
where $\ind{\F(0)\geq\rho}$ is the indicator function of the excursion set,
and in the second to last line, we have taken advantage of the statistical 
homogeneity.

A key question is what information about the statistically homogeneous 
random field $\F$ is contained in volume-based or interfacial-based MTs, 
respectively?
The answer for the volume-based MTs is straightforward since the 
calculation closely follows that of the volume:
\begin{eqnarray}
  w_0^{r,0}(\rho) 
  &=& \lim_{|\Ob|\to\infty} \int_{\Ob} \frac{x^r}{|\Ob|^{1+r/d}} \E\left[\ind{\F(x)\geq\rho}\right] dx\\
  &=& \phi(\rho) \lim_{|\Ob|\to\infty} \int_{\Ob} \frac{x^r}{|\Ob|^{1+r/d}}dx.
  \label{eq:volMT}
\end{eqnarray}
The only information about the Gaussian random field that is contained 
in the densities of the volume-based MTs of arbitrary rank is $\phi(\rho)$.
The scalar volume fraction is simply multiplied by a rescaled MT of the 
observation window.
An intuitive interpretation is that a statistically homogeneous random 
field cannot exhibit any anisotropy in the average distribution of 
``mass''.

This is to be contrasted to the anisotropy of the orientation of the 
interface or curvature, which is quantified by the interfacial-based MT 
densities.
We here derive a general formula that expresses the global average of 
the MT densities by local averages of the functional values and 
gradient.
Therefore, we neglect boundary contributions in the thermodynamic limit 
and apply the coarea formula~\cite{Jensen1998} with $\partial K_{\rho} = 
F^{-1}(\rho)$, where we use the Dirac-delta notation:
\begin{align}
  \label{eq:intMT+1}
  w_{\nu}^{0,s}(\rho)
  &= \lim_{|\Ob|\to\infty} \frac{1}{|\Ob|} \E\left[\int_{\Ob\cap\partial K_{\rho}}n^s(x) H_{\nu-1}(K_{\rho}\cap \Ob,x) dS(x)\right]\\
  &= \lim_{|\Ob|\to\infty} \frac{1}{|\Ob|} \E\left[\int_{\Ob} n^s(x) H_{\nu-1}(K_{\rho}\cap \Ob,x) \|\nabla F(x)\|\delta(F(x)-\rho)dx\right] \\
  &= \lim_{|\Ob|\to\infty} \frac{1}{|\Ob|} \int_{\Ob} \E\Big[n^s(x) H_{\nu-1}(K_{\rho}\cap \Ob,x) \|\nabla F(x)\|\delta(F(x)-\rho)\Big] dx\\
  &= \E\Big[n^s(0) H_{\nu-1}(K_{\rho}\cap \Ob,0) \|\nabla F(0)\|\delta(F(0)-\rho)\Big].
  \label{eq:intMT-1}
\end{align}
Since the gradient of a smooth function is always antiparallel to the 
normal vector on the excursion set, we obtain
\begin{align}
  w_{\nu}^{0,s}(\rho)
  &= (-1)^s\E\Big[H_{\nu-1}(K_{\rho}\cap \Ob,0) \big(\nabla F(0)\big)^s \|\nabla F(0)\|^{-s+1}\delta(F(0)-\rho)\Big].
  \label{eq:intMT_B}
\end{align}
Because of the additivity of the MT, we only need the distributions 
of the functionals value and derivatives at the origin to compute the 
global average of the MT.
This is carried out in the following section for $w_1^{0,s}$ of the 
Gaussian random field. The calculations for $\nu>1$ are technically more 
involved but follow the same principle.

\section{Shape characterization of Gaussian random fields}
\label{sec:GRF}

For a Gaussian random field $G$, we can explicitly evaluate 
Eq.~\eqref{eq:intMT_B} for $\nu=1$ since $G(0)$ and $\nabla G(0)$ are 
stochastically independent (see Sec.~\ref{sec:grf}).
The MT densities $w_1^{0,s}$ of Gaussian random fields in arbitrary 
dimensions are given by:
\begin{eqnarray}
  w_1^{0,s}(\rho)
  &=& (-1)^s\E\Big[\big(\nabla G(0)\big)^s\|\nabla G(0)\|^{-s+1}\delta(G(0)-\rho)\Big]\\
  &=& (-1)^s\int_{\R}\delta(g_0-\rho)\varphi_{G(0)}(g_0)dg_0 \int_{\R^d}\frac{g_1^s}{\|g_1\|^{s-1}}\varphi_{\nabla G(0)}(g_1)dg_1 \\
  &=& (-1)^s\varphi_{G(0)}(\rho) \int_{\R^d}\frac{g_1^s}{\|g_1\|^{s-1}}\varphi_{\nabla G(0)}(g_1)dg_1,
  \label{eq:avgGrad}
\end{eqnarray}

For tensors $w_{1}^{0,2s-1}$ of odd rank, the integrand in Eq.~(\ref{eq:avgGrad}) is anti-symmetric.
Hence, all interfacial MT densities of odd rank vanish:
\begin{eqnarray}
  w_1^{0,2s-1}(\rho) \equiv 0,
  \label{eq:odd-vanishes}
\end{eqnarray}
independent of the covariance functions of the Gaussian random field.
A vanishing of all odd rank tensors implies a centrally symmetric 
distribution of the normal vectors, which is consistent with the fact 
that the covariance function is centrally symmetric.

The interfacial anisotropy is, therefore, exclusively encoded in the 
even rank tensors.
By expressing the remaining integral in Eq.~(\ref{eq:avgGrad}) in 
spherical coordinates, we can carry out the integration of the absolute 
value of the gradient so that only the integration of the direction 
$u\in\mathbb{S}^{d-1}$ of the gradient remains.
For tensors of even rank $s$, we thus obtain:
\begin{eqnarray}
  w_{1}^{0,s}(\rho) &=& \varphi_{G(0)}(\rho)
  \frac{2^{(d-1)/2}\Gamma\big((d+1)/2\big)}{(2\pi)^{d/2}\sqrt{\det C_{\nabla\G(0)}}}
  \int_{\mathbb{S}^{d-1}}\frac{u^s}{\langle u,C_{\nabla\G(0)}^{-1}u\rangle^{(d+1)/2}}du,
  \label{eq:w10s-in-d-dim}
\end{eqnarray}
where $C_{\nabla\G(0)}$ is the covariance matrix of $\nabla\G(0)$ and 
$\Gamma$ is the gamma function.

Remarkably, the MT densities $w_{1}^{0,s}(\rho)$ only trivially depend 
on the threshold $\rho$ via a scalar prefactor $\varphi_{G(0)}(\rho)$.
In other words, the choice of the threshold only changes the trace of 
the tensor $w_{1}^{0,s}(\rho)$, which is, in fact, the specific surface 
$w_1^{0,0}(\rho)$.
In contrast, the actual anisotropy information encoded in the higher 
rank tensors is independent of $\rho$.
This implies that the degree of anisotropy and the preferred orientation 
are intrinsic properties of the Gaussian random field.
Note that this holds for all tensor densities $w_{\nu}^{0,s}(\rho)$ with 
$\nu>0$.

\subsection{Planar Gaussian random fields}

Because the covariance matrix $C_{\nabla\G(0)}$ is symmetric,
a rotation, represented by a rotation matrix $Q$, can diagonalize the covariance matrix.
Using the Cartesian representation, we can write in two dimensions
\begin{eqnarray}
  C_{\nabla\G(0)} = Q
  \left(
  \begin{array}{c c}
    \lambda_1 & 0\\
    0 & \lambda_2
  \end{array}
  \right)
  Q^t.
  \label{eq_grf_diag_cov_def_Q}
\end{eqnarray}
Thus, the integral in Eq.~(\ref{eq:w10s-in-d-dim}) can be easily calculated,
e.g., for the second-rank tensor:
\begin{eqnarray}
  w_1^{0,2}(\rho) &=& \varphi_{G(0)}(\rho) \sqrt{\frac{2}{\pi}}\frac{\sqrt{\lambda_1 \lambda_2}}{\lambda_2 - \lambda _1 } Q
  \left(
  \begin{array}{c c}
    h(\lambda_1,\lambda_2) & 0\\
    0 & -h(\lambda_2,\lambda_1)
  \end{array}
  \right)
  Q^t
\label{eq_grf_w10s_2d}
\end{eqnarray}
with
\begin{eqnarray}
  h(x,y) := \sqrt{x}\left[K\left( 1 - \frac{x}{y} \right)-E\left(1 - \frac{x}{y}\right)\right],
\end{eqnarray}
where $K(k)$ and $E(k)$ are the complete elliptic integral of the first
and second kind, respectively.

\begin{figure}[t]
  \centering
  \includegraphics[width=\linewidth]{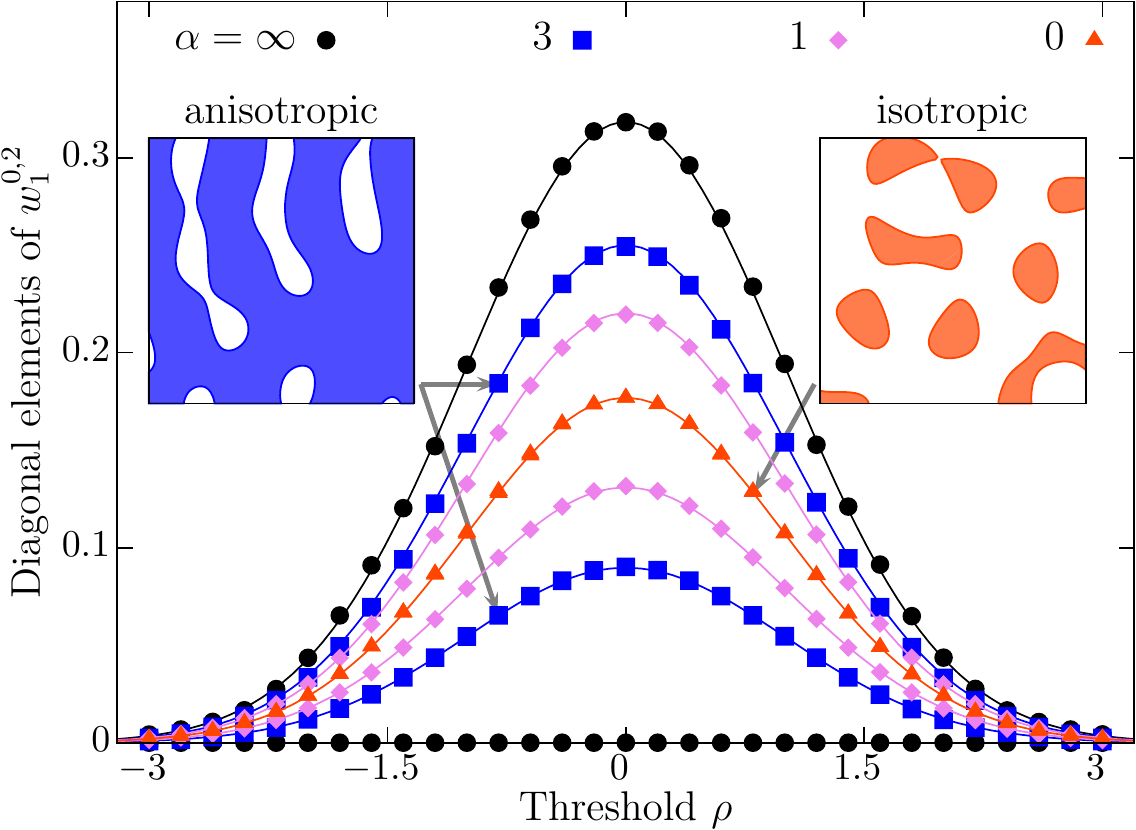}
  \caption{The diagonal elements of the MT density $w_1^{0,2}$ in the 
  Cartesian representation for a Gaussian random field as a function of 
  the threshold $\rho$, see Eq.~(\ref{eq_grf_w10s_2d}), for differently 
  anisotropic Gaussian random wave models.}
  \label{fig_grf_w102}
\end{figure}

To exemplify the behavior of the MT densities, we define a class of 
parametric Gaussian random wave models, in which we can tune the degree 
of anisotropy from an isotropic ($\alpha=0$) to a ``completely anisotropic'' 
model ($\alpha=\infty$).
In the latter case, the excursion sets are effectively one-dimensional 
and consist of parallel slabs with varying thicknesses.
The class of models is defined in detail in \ref{sec:RW}.
There we also describe the simulation procedure and parameters.

Figure~\ref{fig_grf_w102} shows the diagonal elements of the tensor 
$w_1^{0,2}$ in the Cartesian representation as a function of the 
threshold $\rho$.
The lines depict the analytic curves.
The points show numerical estimates from simulated random fields, which 
are in excellent agreement with the exact results.
For the isotropic system, the diagonal elements coincide 
$(w_{1}^{0,2})_{xx}=(w_{1}^{0,2})_{yy}=\frac{1}{2}w_{1}^{0,0}$.
As the degree of anisotropy increases, $(w_{1}^{0,2})_{xx}$ increases 
and $(w_{1}^{0,2})_{yy}$ decreases.
For an effectively one-dimensional model with $\alpha=\infty$, 
$(w_1^{0,2})_{yy}$ vanishes.

\subsection{Additional anisotropy information in higher-rank tensors}

Assuming a diagonal covariance matrix of the gradient, also the fourth 
rank tensor can be calculated straightforwardly (for $\lambda_2 > 0$)
\begin{eqnarray}
  (w_1^{0,4})_{xxxx} &=& \varphi_{G(0)}(\rho) \sqrt{\frac{2}{\pi}} \frac{\lambda_1\left((\lambda_1+\lambda_2)E( 1 - \frac{\lambda_1}{\lambda_2} )-2\lambda_1 K( 1 - \frac{\lambda_1}{\lambda_2} )\right)}{\lambda_2\sqrt{\lambda_2}(1-\frac{\lambda_1}{\lambda_2})^2}, \\
    (w_1^{0,4})_{yyyy} &=&  \frac{\lambda_2}{\lambda_1}(w_1^{0,4})_{xxxx},\\
    (w_1^{0,4})_{xxyy} &=& (w_1^{0,2})_{xx} - (w_1^{0,4})_{xxxx},\\
    (w_1^{0,4})_{xyyy} &=& (w_1^{0,4})_{xxxy} = 0.
\label{eq_grf_w104}
\end{eqnarray}
In the Cartesian representation, it is 
non-trivial to check whether the fourth-rank tensor contains additional 
anisotropy information compared to the second-rank tensor.
We, therefore, switch to the irreducible representation, which is 
explained in detail in \ref{sec:irr}.

There we derive from the irreducible representation with respect to 
$\mathrm{SO}(2)$ scalar anisotropy indices $q_s\in[0,1]$ that 
characterize the degree of anisotropy.
If $q_s>0$, anisotropy is detected by the tensor of rank $s$ that is not 
captured by tensors of another rank.
Explicit expressions for the irreducible anisotropy indices $q_2$ and 
$q_4$ can be derived for planar Gaussian random fields by inserting 
Eqs.~(\ref{eq_grf_w10s_2d}) and (\ref{eq_grf_w104}) into 
Eqs.~(\ref{eq_cmt_q2}) and (\ref{eq_cmt_q4}).

\begin{figure}[t]
  \hspace{3.4cm}
  \begin{minipage}[t]{0.2\linewidth}
    \centering
    Rank 2
  \end{minipage}%
  \hspace{4.8cm}
  \begin{minipage}[t]{0.2\linewidth}
    \centering
    Rank 4
  \end{minipage}

  \vspace{0.2cm}

  \includegraphics[width=0.48\linewidth]{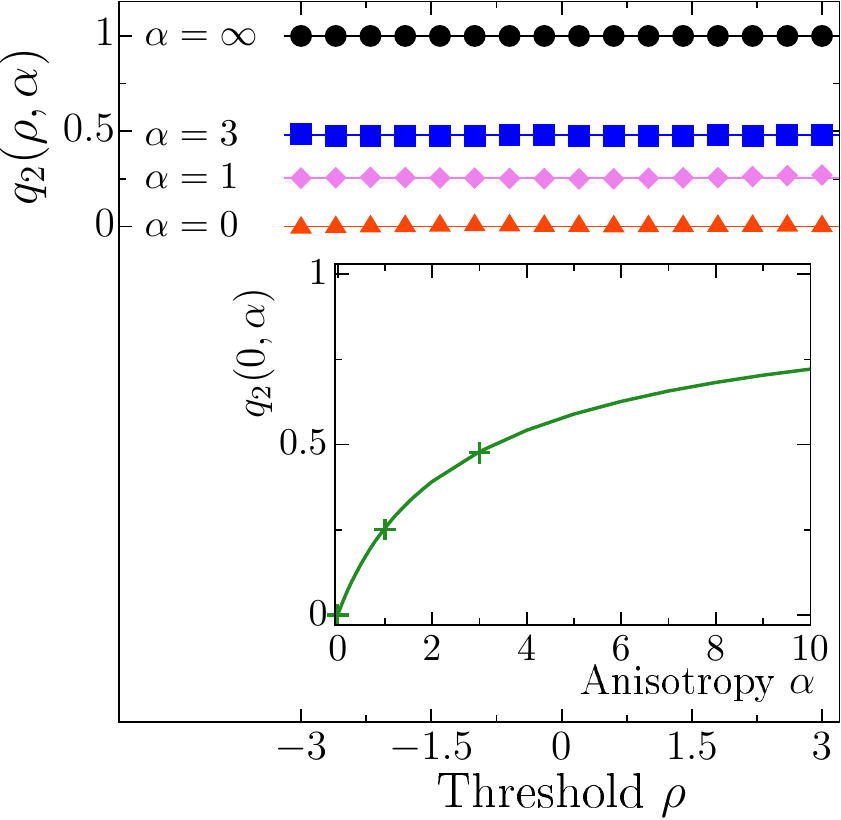}
  \hfill
  \includegraphics[width=0.48\linewidth]{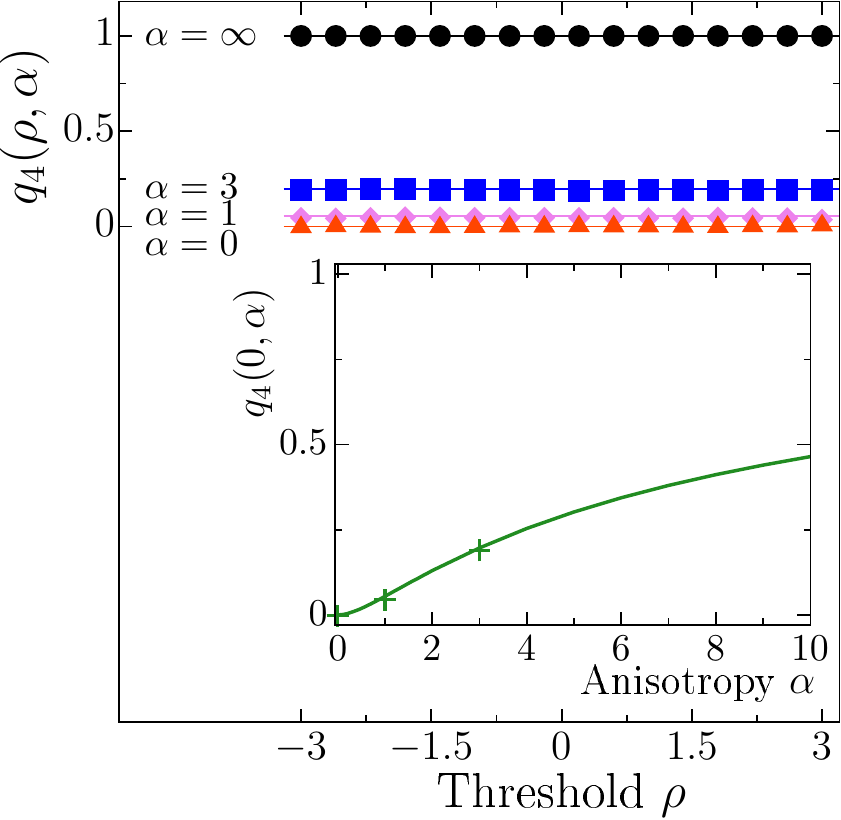}
  \caption{Anisotropy indices of rank two and four for differently anisotropic Gaussian random wave models:
    The indices $q_2$ and $q_4$ are the ratio of the absolute value of the irreducible representations of rank 2 and rank 0 or rank 4 and rank 0, respectively, see Eqs.~(\ref{eq_cmt_q2}) and (\ref{eq_cmt_q4}).
    The indices as a function of the threshold $\rho$ are constant, that is, independent of $\rho$.
    The insets show the indices as a function of the anisotropy parameter $\alpha$ of the models.
    Because $q_4>0$, the fourth rank tensor $w_1^{0,4}$ contains nonredundant anisotropy information compared to the second-rank tensor.}
  \label{fig_grf_anisotropy_indices}
\end{figure}

The resulting data for our Gaussian random waves models are plotted in 
Fig.~\ref{fig_grf_anisotropy_indices}.
As before, the lines depict the analytic curves, and the points 
represent the simulation results, which are again in excellent agreement 
with the formulas.
Since the degree of anisotropy is, as explained above, an 
inherent property of the Gaussian random field, the anisotropy indices 
are independent of the threshold $\rho$.
Importantly, we find that $q_4>0$ for anisotropic Gaussian random 
fields, i.e., the tensor of rank four captures additional anisotropy 
information that is not contained in the second-rank tensor.

\subsection{Higher-rank tensors predicted from second-rank tensors}

What this example demonstrates is that higher-rank tensors of Gaussian 
random fields will generally contain additional anisotropy information.
The next question is: what kind of additional information is contained 
in the higher-rank tensors?

\begin{figure}[t]
  \flushright
  \includegraphics[width=\linewidth]{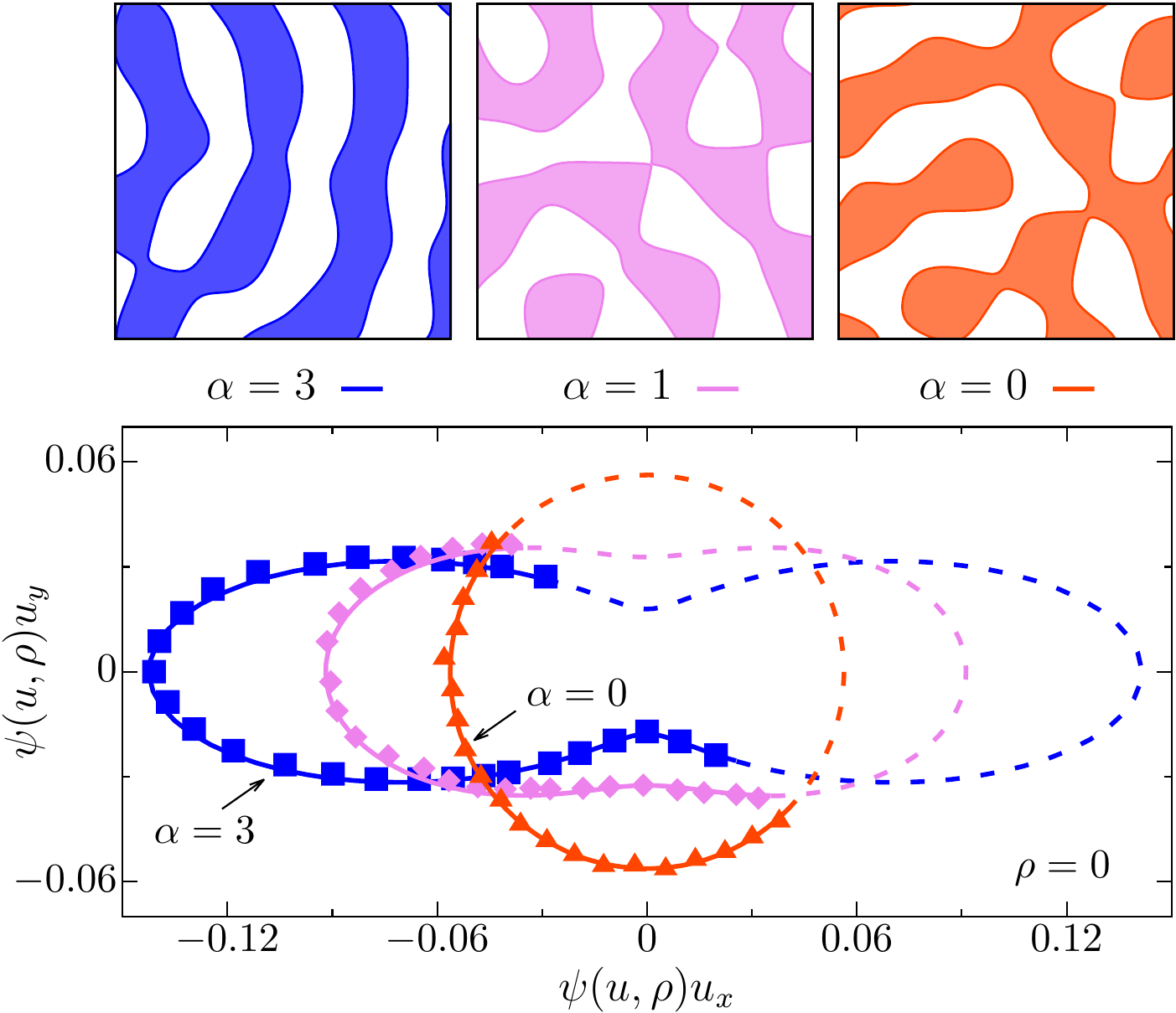}
  \caption{The EGI density $\egi(u,\rho)$ from Eq.~(\ref{eq_grf_egi})
    is shown for Gaussian random wave models with different degrees of 
    anisotropy (from isotropic $\alpha=0$ to strongly anisotropic 
    $\alpha=3$).
    At the top, samples of the excursion sets are depicted. The main 
    figure is a polar plot of the EGI densities.
    The solid lines show the analytic curves, which are compared to the 
    direct numerical estimates of $\egi(u,\rho)$ represented by the 
    marks, which are in very good agreement with the prediction from the 
    formula.
    The dashed lines correspond to estimates of the EGI density that are 
    based only on the measurement of $w_1^{0,2}$ (via the 
    method of moments).
    This estimate is in even better agreement with the analytic curves. 
    (Note that the polar plot of $\egi(u,\rho)$ is point symmetric with 
    respect to the origin.)}
  \label{fig_grf_egi}
\end{figure}

To answer this question, we express the MTs $W_1^{0,s}(K)$ as moment tensors of the 
distribution of its normal vectors, or more precisely, of its extended 
Gaussian image $\egiK$, as in \ref{sec:irr}:
\begin{eqnarray}
  W_1^{0,s}(K) = \int_{\mathbb{S}^{d-1}} u^s \egiK(u) du \text{ with } \egiK(u) &:= \int_{\partial K} \delta(n(x)-u)dS(x).
  \label{eq:mainegi}
\end{eqnarray}
The MT densities follow from a corresponding density $\psi(u,\rho)$ of the EGI:
\begin{eqnarray}
  \egi(u,\rho) &:=& \lim_{|\Ob|\to\infty} \frac{\E[\Psi_{K_{\rho}\cap\Ob}(u)]}{|\Ob|} 
\end{eqnarray}
where $u\in\mathbb{S}^{d-1}$.
In analogy to Eqs.~\eqref{eq:intMT+1}--\eqref{eq:intMT-1}, we derive for 
a general random field $F$ (subject to our assumptions in Sec.~\ref{sec:RFMT}):
\begin{eqnarray}
  \egi(u,\rho)
  &=& \lim_{|\Ob|\to\infty} \frac{1}{|\Ob|}\E\left[\int_{\Ob\cap\partial K} \delta(n(x)-u)dS(x)\right]\\
  &=& \lim_{|\Ob|\to\infty} \frac{1}{|\Ob|}\E\left[\int_{\Ob} \delta(n(x)-u)\|\nabla F(x)\|\delta(F(x)-\rho)dx\right]\\
  &=& \lim_{|\Ob|\to\infty} \frac{1}{|\Ob|}\int_{\Ob} \E\big[\delta(n(x)-u)\|\nabla F(x)\|\delta(F(x)-\rho)\big]dx\\
  &=& \E\big[\delta(n(0)-u)\|\nabla F(0)\|\delta(F(0)-\rho)\big].
  \label{eq_egi_general}
\end{eqnarray}
For a Gaussian random field $G$, 
we obtain more explicitly---in analogy to the derivation of Eq.~(\ref{eq:avgGrad}):
\begin{eqnarray}
  \egi(u,\rho)
  &=& \E\Big[\delta\left(\frac{\nabla G(0)}{\|\nabla G(0)\|}+u\right)\|\nabla G(0)\|\Big]\E\Big[\delta\big(G(0)-\rho\big)\Big]\\
  &=& \varphi_{G(0)}(\rho)\int_{\R^d}\delta\left(\frac{g_1}{\|g_1\|}+u\right)\|g_1\|\varphi_{\nabla G(0)}(g_1)dg_1
\end{eqnarray}
For a Gaussian random field, the interfacial anisotropy depends on 
the threshold only via a scalar prefactor, as discussed above.
To evaluate the remaining integral, we separate the integral over the 
absolute value $\gamma_1:=\|g_1\|$ and direction 
$n_1:={g_1}/{\|g_1\|}$ of the gradient:
\begin{eqnarray}
  \egi(u,\rho)
  &=& \varphi_{G(0)}(\rho) \int_{\R} \gamma_1^{d}\int_{\mathbb{S}^{d-1}} \delta\left(n_1+u\right) \varphi_{\nabla G(0)}(\gamma_1 n_1) dn_1 d\gamma_1\\
  &=& \varphi_{G(0)}(\rho) \int_{\R} \gamma_1^{d} \varphi_{\nabla G(0)}(\gamma_1 u) d\gamma_1,
\end{eqnarray}
where we have used the symmetry of $\varphi_{\nabla G(0)}$ in the last 
equation.
Inserting the Gaussian probability density function of the gradient 
yields the explicit expression for the EGI density in any dimension $d$:
\begin{eqnarray}
\egi(u,\rho) &=& \varphi_{G(0)}(\rho)
  \frac{2^{(d-1)/2}\Gamma\big((d+1)/2\big)}{(2\pi)^{d/2}\sqrt{\det C_{\nabla\G(0)}}}
  \times
  \frac{1}{\langle u,C_{\nabla\G(0)}^{-1}u\rangle^{(d+1)/2}}.
  \label{eq_grf_egi}
\end{eqnarray}
Exemplary curves of $\egi(u,\rho)$ are shown in Fig.~\ref{fig_grf_egi} 
for our parametric Gaussian random wave model.
For the isotropic system, the polar plot forms a circle.
For the anisotropic systems, the curves are dumbbell-shaped.
The more anisotropic the field, the more anisotropic the polar plot is, 
i.e., more normal vectors are pointing in the horizontal than in the 
vertical direction.

The MT densities of arbitrary rank easily follow from the EGI density
because the averaging of the MT and the limit are linear operations.
Therefore, Eq.~(\ref{eq:mainegi}) can be directly transferred to the 
densities:
\begin{eqnarray}
  w_{1}^{0,s}(\rho) &=& \int_{\mathbb{S}^{d-1}} {u}^s\egi(u,\rho)du.
  \label{eq_grf_w10s_from_egi}
\end{eqnarray}
Note the agreement with Eq.~(\ref{eq:w10s-in-d-dim}).

Now we can return to the question of what kind of additional information is 
contained in the higher-rank tensors?
Equation~(\ref{eq_grf_egi}) shows that the information about the 
Gaussian random field that is encoded in the EGI density and thus in all 
MT densities is the covariance matrix $C_{\nabla G(0)}$ of the gradient 
at the origin (and the variance of the functional value, which 
is determined by the choice of units).
From the derivation, we can conclude that the underlying reason for this 
parametrization of $\egi(u,\rho)$ via $C_{\nabla G(0)}$ is the 
stationarity of the random field together with the additivity of the MT.
For that reason, the EGI density only depends on the distributions of 
the functional value and the gradient of the Gaussian random field at 
the origin, see Eq.~(\ref{eq_egi_general}).
Because both are centered Gaussian distributions, they are entirely 
determined by their variances and covariances.

The essential information about the Gaussian random field, which is 
encoded in the tensors, is thus the covariance matrix 
$C_{\nabla G(0)}$.
Because it is a symmetric second-rank tensor, its elements can actually 
be estimated from the second-rank MT density $w_1^{0,2}$.
Note that this can already be deduced from Eq.~(\ref{eq:w10s-in-d-dim}).
Still, the EGI density offers a more transparent presentation, and we 
employ it in the following to demonstrate that by measuring $w_1^{0,2}$, we 
can robustly estimate the EGI density and thus all higher-rank tensors 
of the Gaussian random field.

We, therefore, apply the method of moments exemplarily in two 
dimensions.
First, we numerically solve Eq.~(\ref{eq_grf_w10s_2d}) for the 
eigenvalues $\lambda_1$ and $\lambda_2$ of $C_{\nabla G(0)}$.
Then, we numerically estimate $w_1^{0,2}$ based on $3000$ samples of 
excursion sets of the Gaussian random wave models from 
Fig.~\ref{fig_grf_egi} at threshold $\rho=0$.
We then use the eigenvectors of $w_1^{0,2}$ determine the rotation 
matrix $Q$.
Using the ratio of the eigenvalues, we then estimate the ratio of 
$\lambda_1$ and $\lambda_2$.
Finally, we use the trace of $w_1^{0,2}$ to estimate the absolute values 
of $\lambda_1$ and $\lambda_2$.

To confirm the outcome, we insert the estimated $\hat{Q}$, 
$\hat{\lambda_1}$, and $\hat{\lambda_2}$ in Eq.~(\ref{eq_grf_egi}) and 
plot the resulting curve for $\egi(u,\rho)$ in Fig.~\ref{fig_grf_egi} 
(dashed lines).
It is in excellent agreement with the exact analytic curve of the EGI 
density (solid line).
The statistical errors are distinctly smaller than for the points that
represent the direct estimate of $e(u,\rho)$ for some directions $u$ 
(based on a data set twice as large).
The second-rank MT density suffices for a robust and accurate estimate 
of the EGI density and, thus, all higher-rank tensors.

At first, it might sound like a contradiction that, on the one hand, higher-rank tensors 
contain additional information, as we discovered in the previous 
section, but that, on the other hand, we have now predicted this additional information from 
the second-rank tensor.
The key difference is that in the previous section, we did not have to 
assume that the random field is Gaussian to compute the anisotropy 
indices $q_s$ and thus extract additional information.
In contrast, our prediction of the higher-rank tensors is only valid for 
Gaussian random fields.

Let us illustrate this finding with an elucidating geometrical analogy: the 
anisotropy of a rectangle.
If we only know the second-rank tensor of a convex body, we cannot 
reconstruct its shape without further prior knowledge.
The infinite series of interfacial MTs $W_1^{0,s}$ is needed to 
reconstruct its shape, which is---in this case---a rectangle.
The higher-rank tensors thus obviously contain additional information.
However, if we assume that the shape is a rectangle, then the 
second-rank tensor suffices to estimate the perimeter and the aspect 
ratio of the rectangle.
This determines its shape, and thus, we can predict all higher-rank tensors.
Interestingly, we can use these relations to test the assumption by 
comparing our predictions with the measured values.

In the same way, the relation between tensors of different rank that we 
found for Gaussian random fields can be used to detect non-Gaussianities 
in anisotropic random fields.
Because the relation does not depend on the details of the model, that 
is, on the correlation function of the Gaussian random field, it would 
be ideally suitable for a null hypothesis test.
We discuss this outlook in the following section.

\section{Conclusion and outlook to a robust detection of non-Gaussianities}
\label{sec:conclusion}

We have shown that MTs detect and quantify anisotropy in random fields 
robustly and comprehensively; see, e.g., Fig.~\ref{fig_grf_w102}.
Our simulations demonstrate that the asymptotic results are robust 
against noise and finite-size effects.
Therefore, we are convinced that our shape analysis can be useful for 
experiments and applications in physics, where the Gaussian random field 
is already a common model; e.g., for rough 
surfaces~\cite{filliger_3d_2012, lessel_impact_2013}, porous 
media~\cite{teubner_level_1991, RobertsTorquato1999}, or as solutions to 
stochastic partial differential equations like the heat 
equation~\cite{Xiao2009}, but also for other fields of science and 
technology, including image processing, geostatistics, or spatial 
statistics.

\paragraph{Shape information encoded in MTs}
Importantly, we have carefully discussed the different geometrical 
information contained in tensors of different order (e.g., volume-based 
versus interfacial based MTs) and of different rank.
For quite general random fields $\f$ (that are ergodic, sufficiently 
smooth and integrable), the volume tensors only contain the volume of 
the excursion set and shape information about the observation window, 
see Eq.~(\ref{eq:volMT}).
For the interfacial tensors, the global average of the MTs can be 
expressed in the thermodynamic limit by a local average of the 
functional value and the gradient at the origin, see 
Eq.~(\ref{eq:intMT_B}).
They are suitable to characterize the global anisotropy of ergodic 
random fields since they quantify the intrinsic anisotropy of the 
Gaussian random field (independent of the threshold).

Due to the symmetry of the Gaussian random field, all MT tensor 
densities $w_1^{0,2s-1}$ of odd rank vanish.
For the even rank tensor densities, we derive an explicit integral 
expression for arbitrary dimensions, see Eq.~(\ref{eq:w10s-in-d-dim}).
In two dimensions, we derive the explicit expression in 
Eq.~(\ref{eq_grf_w10s_2d}).
Comparing the analytic results to simulations, see 
Figure~\ref{fig_grf_w102}, we show how the MT densities detect 
anisotropy in the Gaussian random field and quantify the degree of 
anisotropy.
We demonstrate this for a parametric model with a tunable degree of 
anisotropy, as described in \ref{sec:RW}.

The irreducible representation of MTs reveals that tensors of different 
rank contain additional information about the anisotropy of the level 
sets, see Fig.~\ref{fig_grf_anisotropy_indices};
for an introduction to the irreducible representation, see 
\ref{sec:irr}.
The more tensors we determine, the more accurately the level sets are
characterized without prior knowledge about the random field.

Perhaps surprisingly, we have shown that if we know the random field is 
Gaussian, then this additional information can be predicted from the 
second-rank tensor.
This is because the essential information about the Gaussian random 
field encoded in the Minkowski tensors is a second rank tensor, the 
covariance matrix of the gradient at the origin, see 
Eq.~(\ref{eq_grf_egi}).
The second-rank tensor $w_1^{0,2}$ thus accesses all information from 
the Gaussian random field that is needed to determine the higher-rank 
tensors $w_1^{0,s}$ (once we know that the random field is Gaussian).

Based on a measurement of $w_1^{0,2}$, we can estimate the EGI density, 
see Fig.~\ref{fig_grf_egi}, and thus all tensor densities $w_1^{0,s}$, 
see Eq.~(\ref{eq_grf_w10s_from_egi}).
Our simulations have demonstrated that a moderately large collection of 
samples with a relatively small system size is sufficient for a robust 
and accurate estimate.

\paragraph{Basis for a robust null hypothesis test}
Our relation between tensors of different rank
provides the basis for a null hypothesis test detecting 
non-Gaussianities in anisotropic random spatial structures.
First, the second-rank tensor is measured.
Then its prediction of, e.g., the fourth-rank tensor can be 
compared to the measured value of the tensor of rank four.
A statistically significant deviation of the prediction from the 
measurement would indicate a non-Gaussian contribution.
Such a test could, for example, be constructed similar to the test on 
complete spatial randomness using Minkowski functionals in 
\cite{ebner_goodness--fit_2018,klatt_detecting_2019}.

A major advantage of such a test is the complete independence from the 
models of the Gaussian random field; essentially, we only assume 
integrability and smoothness as mentioned above.
For a rigorous hypothesis test, the covariances of MTs and central limit 
theorems are needed, which were recently proven for the MFs in the 
thermodynamic 
limit~\cite{paper:Mueller,mueller2018,paper:KratzVadlamani},
as well as for an average of interfacial MTs over all 
thresholds~\cite{mueller2018}.
The exact values of the covariances will probably depend on the details 
of the model.
Nevertheless, we expect that a null hypothesis test that detects 
non-Gaussianities in anisotropic random fields using MTs would be robust 
against minor changes in the model because the relation 
between the mean values of MTs is universal.

\ack

We are grateful to Maria Schlecht and Daniel G\"oring for their 
preliminary work on the shape of Gaussian random fields.
We thank Dennis M\"uller for our insightful discussions.
We thank Anne Estrade and Julie Fournier for stimulating discussions and 
Mark Dennis for pointing out the random waves model.
M.A.K.~and K.M.~acknowledge support by the Deutsche 
Forschungsgemeinschaft (DFG) through the SPP 2265, under grant number ME 
1361/16-1, and by the Volkswagenstiftung via the ``Experiment'' Project 
``Finite Projective Geometry.''

\appendix
\section{Parametric Gaussian random wave model}
\label{sec:RW}

For our simulations, we use a parametric model of a Gaussian random 
field, where we can tune the degree of anisotropy and preferred 
orientation.
Our model belongs to the class of Gaussian random wave 
models~\cite{berk_scattering_1991, berry_regular_1977, 
dennis_nodal_2007}, which is an exemplary class of Gaussian random 
fields.
These models are especially important in physics; for example, Gaussian 
random waves solve the time-independent Helmholtz wave 
equation~\cite{dennis_nodal_2007} and are physically well-justified 
models for chaotic quantum systems~\cite{BaeckerSchubert2002, 
UrbinaRichter2003}.
Moreover, they are connected to optical speckle 
patterns~\cite{Goodman2007speckle, Goodman1985statistical}.
In contrast to the typical isotropic models, our model has, as mentioned 
above, a tunable degree of anisotropy.
Our parametric model was first defined in 
\cite{klatt_morphometry_2016} and has also been studied in 
\cite{estrade_anisotropic_2020}.

The basic idea is to superimpose plane waves with random phases $\eta_i$ 
and a random direction $k_i$ of the wave vector:
\begin{eqnarray}
  \g(x)&=&\sqrt{\frac{2}{N_{w}}}\sum_{i=1}^{N_{w}}\cos(k_ix+\eta_i), 
  \label{eq_grf_grwm_def}
\end{eqnarray}
where the random phases $\eta_i$ are uniformly distributed on $[0,2\pi)$ 
and $\|k_i\|\equiv1$.
We note in passing that, by our choice that $\|k_i\| \equiv 1$, our 
Gaussian models are stealthy hyperuniform, i.e., they anomalously 
suppress density fluctuations on large scales~\cite{torquato_local_2003, 
ma_random_2017, torquato_hyperuniform_2018}. Variants of our model can 
be used to construct Gaussian random fields that are only anisotropic in 
specific directions~\cite{torquato_hyperuniformity_2016}.

\begin{figure}[t]
  \flushright
  \includegraphics[width=0.47\linewidth]{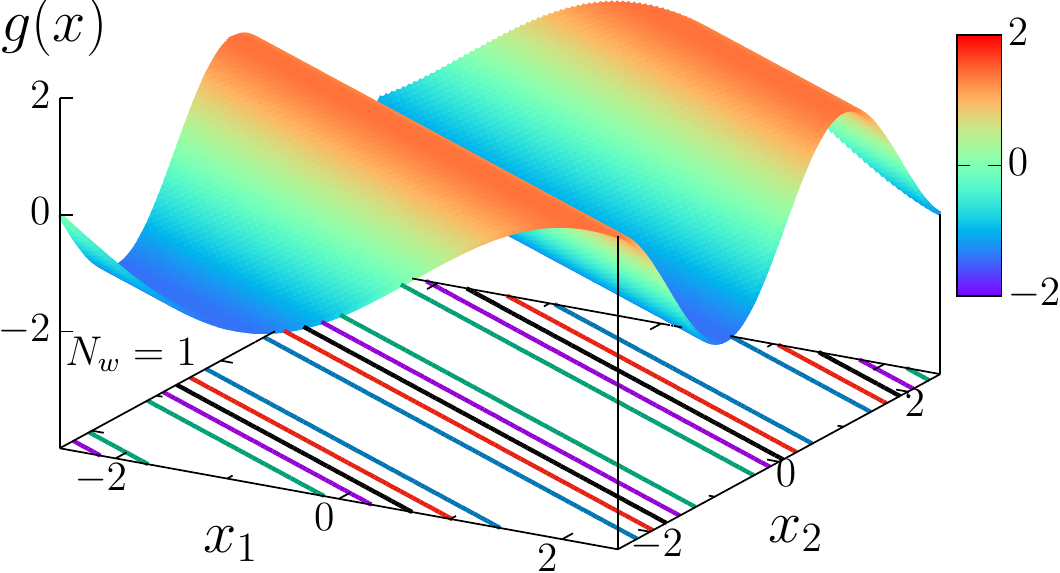}
  \hfill
  \includegraphics[width=0.47\linewidth]{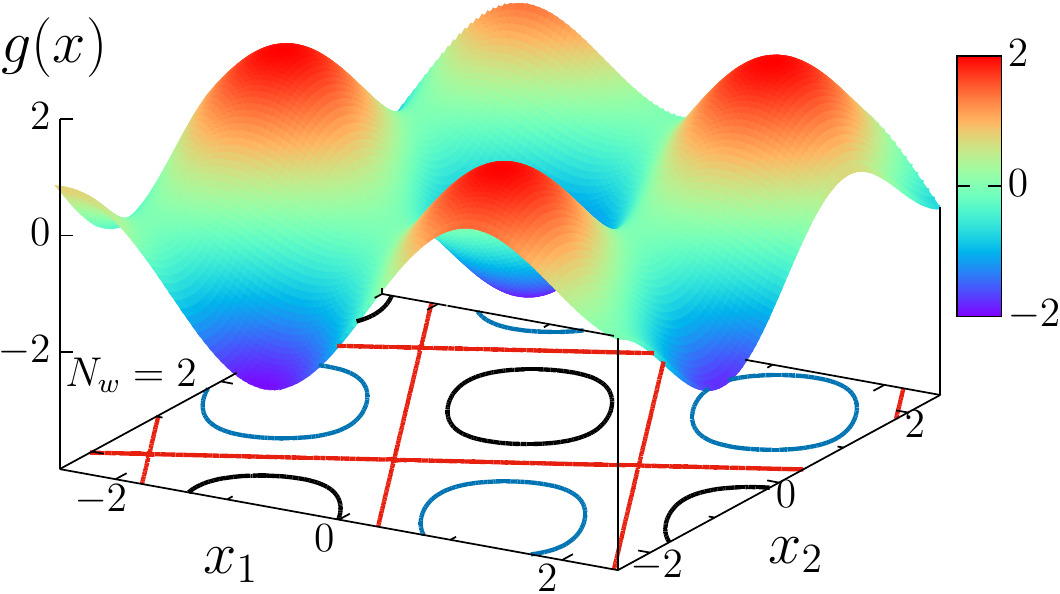}\\
  \vspace{0.2cm}
  \includegraphics[width=0.47\linewidth]{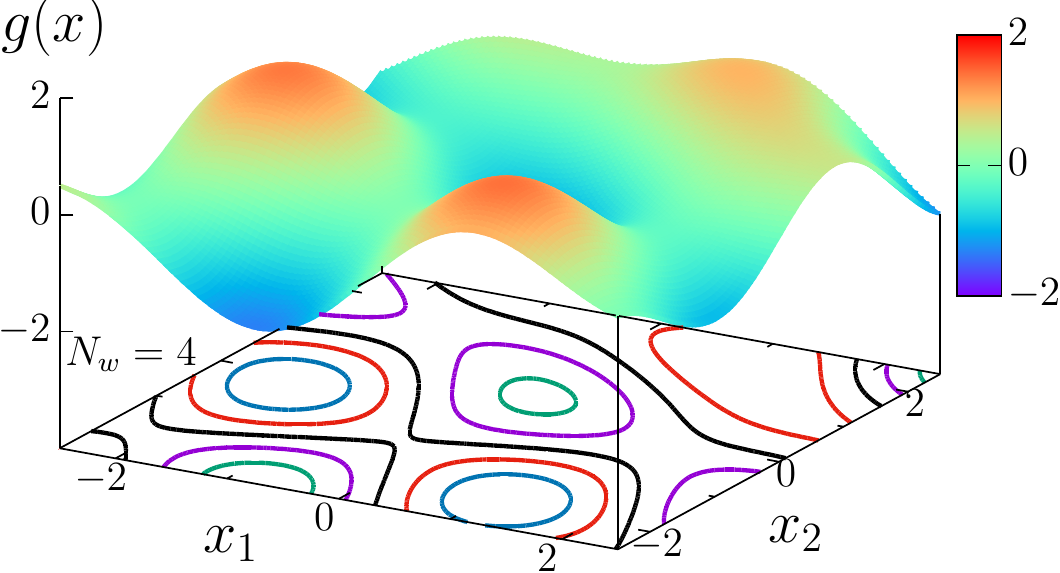}
  \hfill
  \includegraphics[width=0.47\linewidth]{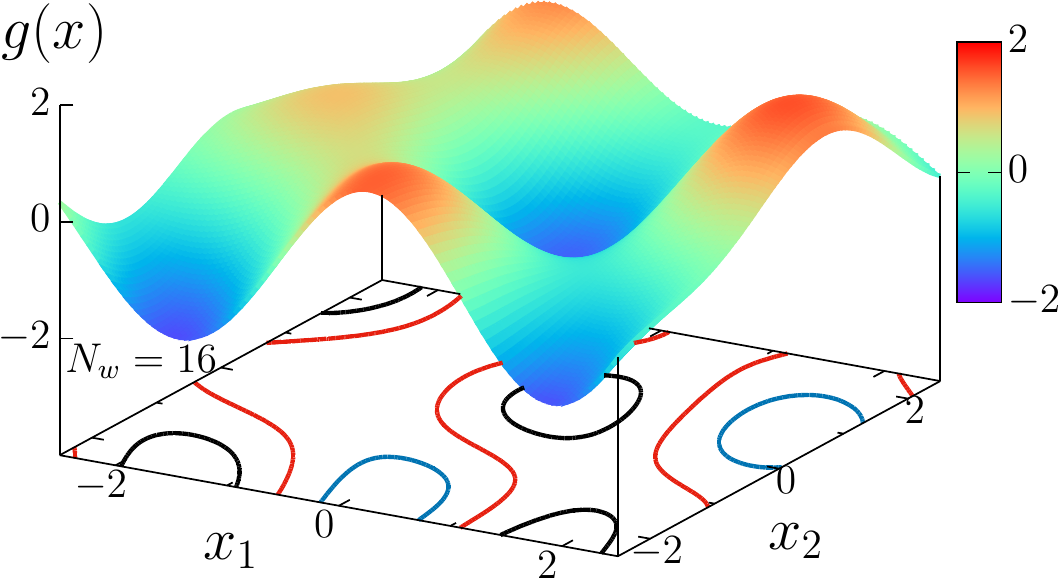}
  \caption{Gaussian random waves, see Eq.~(\ref{eq_grf_grwm_def}), the 
  plots show the superposition of $N_w=1,2,4,16$ plain waves with random 
  phases and orientation. It quickly converges to a Gaussian random 
  field.}
  \label{fig:grw}
\end{figure}

At each position $x$, the functional values of the random waves are 
independently and identically distributed random variables.
Because of the central limit theorem, the superposition is, therefore, 
approximately a Gaussian random field for a large number $N_{w}$ of 
random waves, where in our simulation, $N_w=100$.
Figure~\ref{fig:grw} shows the stepwise construction of our Gaussian 
random wave model for the sample in Fig.~\ref{fig_grf_sample}.
The distribution of $k_i$ on $\mathbb{S}^{d-1}$ controls the anisotropy 
of the system.
The model is isotropic for a uniform distribution on $\mathbb{S}^{d-1}$.
It is anisotropic if there is an orientation bias in the distribution of 
$k_i$.

Here we describe a two-dimensional model, where we parametrize the 
direction $k_i\in\mathbb{S}^1$ by the angle $\theta_i$ between $k_i$ and 
the $x$-axis.
We choose for the probability density of $\theta$ (similar to the 
anisotropic distributions in~\cite{schroder-turk_minkowski_2011, 
horrmann_minkowski_2014, klatt_anisotropy_2017, 
klatt_mean-intercept_2017}):
\begin{eqnarray}
  f_{\alpha}(\theta) &=& Z_{\alpha} \cos^{\alpha}(\theta)
  \label{orientation_distribution}
\end{eqnarray}
with $Z_{\alpha}=\Gamma(1+\alpha/2)/(\sqrt{\pi} \Gamma(1/2+\alpha/2))$ 
with $\theta\in (-\pi/2,\pi/2]$.
The parameter $\alpha$ varies the anisotropy from perfect alignment 
($\alpha=\infty$) to an isotropic distribution ($\alpha=0$).
Interestingly, for fixed $\alpha=1$, the anisotropic Gaussian random 
waves can model polarized light fields~\cite{de_angelis_spatial_2016}.

The power spectral density $\psd(q)$ follows directly from the 
multivariate probability density function of $k_i$:
\begin{eqnarray}
  \psd(q) = 2\pi^2 \delta(\|q\|-1) Z_{\alpha} |\cos(\theta)|^{\alpha},
  \label{eq_grwm_sigma_tilde}
\end{eqnarray}
where $\theta\in[-\pi,\pi)$ is the angle between $q$ and the $x$-axis.
In contrast to the orientation distribution in 
Eq.~(\ref{orientation_distribution}), 
$Z_{\alpha}|\cos(\theta)|^{\alpha}$ is defined for all directions.

The eigenvalues $\lambda_1$ and $\lambda_2$ of the covariance matrix 
$\Cov[\nabla\g(0)]$ of the gradient of $\g$ at the origin are
\begin{eqnarray}
 \lambda_1 &=& \frac{1+\alpha}{2+\alpha}\quad\mathrm{and}\quad 
 \lambda_2 = \frac{1}{2+\alpha},
\end{eqnarray}
and depend as expected on the anisotropy of the model.
In the limit $\alpha\rightarrow\infty$, all wave vectors are aligned in 
the same direction, i.e., all wavefronts are parallel to each other.
The model becomes an effectively one-dimensional random field that is 
simply extended perpendicular to the wavefronts. Accordingly for 
$\alpha\to\infty$,
$\lambda_1 {\to} 
1$ and $\lambda_2 {\to} 0$.
In the isotropic case, $\alpha=0$, the eigenvalues coincide 
$\lambda_1=\lambda_2={1}/{2}$.

We simulate for each parameter $\alpha=0,1,3$, $1000$ samples within the 
observation window $[0,20)^2$ for Figs.~\ref{fig_grf_w102} and 
\ref{fig_grf_anisotropy_indices}, and $2000$ samples for 
Fig.~\ref{fig_grf_egi}.
In our simulation, we first sample the parameters of the function 
$\g(x)$ in Eq.~(\ref{eq_grf_grwm_def}).
At each threshold, we approximate the level set by a polygon, using the 
software \textsc{gnuplot}, for which the MTs and EGI can 
quickly be evaluated using a \textsc{python} 
script.
To sample the random numbers, we use the MT19937 
generator, which is known as the 
``Mersenne Twister'', implemented in the \textsc{Boost 
Libraries}.

\section{Irreducible representation of Minkowski tensors}
\label{sec:irr}

An irreducible representation of Minkowski tensors in two dimensions 
provides convenient access to anisotropy information of high-rank 
tensors.
Importantly, such an irreducible representation avoids redundant 
information in tensors of different rank.
In that sense, the irreducible representation of a tensor of rank $s$ is 
the anisotropy information that is exclusive to this rank.
This is in contrast to the Cartesian representation, where the definition 
is intuitive but where the same information can be contained in tensors 
of different rank.

We decompose the Minkowski tensors into irreducible tensors with respect 
to the rotation group $\mathrm{SO}(d)$; see \cite{Kanatani:1984} for a 
discussion of irreducible representations of moment tensors of the 
directional distributions and \cite{AleskerBernigSchuster2011} for a 
general theory of an irreducible decomposition of the space of 
translational-invariant and continuous valuations.
For three dimensions, an explicit irreducible representation and 
rotational invariants of Minkowski tensors are presented in 
\cite{Kapfer2011, kapfer_jammed_2012, MickelEtAl2013}.
Here, the two-dimensional case of \textit{irreducible Minkowski tensors} 
(IMTs) for the interfacial tensors $W_1^{0,s}$ for a convex body $K$ is 
discussed; see also \cite{klatt_mean-intercept_2017, 
collischon_tracking_2021}.
Note that the IMTs are equivalent to the definition of the harmonic 
intrinsic volumes defined in \cite{Hoermann2015}.

Here we provide an intuitive approach to the IMTs based on an 
interpretation of $W_1^{0,s}$ as being essentially moment tensors of the 
normals on the interface of a domain~$K$:
\begin{eqnarray}
  W_1^{0,s}(K)
  &=&  \int_{\partial K}n^s(x)dS(x)\\
  &=&  \int_{\partial K}\int_{\mathbb{S}^{d-1}}u^s\delta(n(x)-u)dudS(x)\\
  &=&  \int_{\mathbb{S}^{d-1}} u^s \int_{\partial K}  \delta(n(x)-u) dS(x)du\\
  &=&  \int_{\mathbb{S}^{d-1}} u^s \egiK(u) du,
\label{eq_cmt_w10s_via_egi}
\end{eqnarray}
where
\begin{eqnarray}
  \egiK(u) &:=& \int_{\partial K} \delta(n(x)-u)dS(x)
\end{eqnarray}
is the so-called \textit{extended Gaussian image} (EGI)~\cite{horn_extended_1984}.
It is defined on the unit sphere (or circle in $d=2$) $\mathbb{S}^{d-1}$ and is proportional to the probability 
density function of the normal vectors, but its normalization is equal 
to the surface area (or perimeter in $d=2$) $W_1(K)=\int_{\mathbb{S}^{1}} \egiK(u) du$.

In two dimensions, the direction $u\in\mathbb{S}^{1}$ can be identified 
by the angle $\theta$ between $u$ and the $x$-axis.
Then $\egiK(\theta)$ is a periodic function on $[0,2\pi)$ and can be 
represented by a Fourier series
\begin{eqnarray}
  \egiK(\theta)&=&  \sum_{s=-\infty}^{\infty}\egiC_s(K)\exponent{is\theta}, \mathrm{ with}\\
  \psi_s(K)    &:=& \frac{1}{2\pi}\int\limits_0^{2\pi}\egiK(\theta) \exponent{-is\theta} d\theta,
  \label{eq_cmt_egi_def}
\end{eqnarray}
so that $\egiC_{-s}(K) = \egiC_{s}^*(K)$.
Because $\exponent{is\theta}$ with $s\in\mathbb{Z}$ are the 
irreducible representations of $\mathrm{SO}(2)$,
the Fourier coefficients $\egiC_s(K)$ of the EGI are the IMTs, that is, the 
irreducible representations of the translation invariant interfacial 
tensors $W_1^{0,s}(K)$.
For a rotation of $K$ by an angle $\beta$, i.e., $\egiK(\theta)\rightarrow\egiK(\theta-\beta)$, we obtain for the IMTs 
 the simple transformations
\begin{eqnarray}
  \egiC_s(K) &\rightarrow&\; \exponent{-is\beta}\egiC_s(K).
\end{eqnarray}
Hence, the phase of the complex number $\egiC_s(K)$ contains information 
about the preferred direction.
Its absolute value $|\egiC_s(K)|$ is a scalar index (or rotational invariant).
For $s=0$, $\egiC_0(K)={W_1}/(2\pi)$, and $\egiC_{\pm1}(K)=0$ for a closed 
body $K$.
For $|s|\geq2$, $|\egiC_s(K)|$ quantifies the degree of anisotropy.
Hence, we define the anisotropy indices 
\begin{eqnarray}
  q_s(K) &:=& \frac{|\egiC_s(K)|}{\egiC_0(K)}  \in [0,1],
  \label{eq_cmt_anisotropy_index}
\end{eqnarray}
which are unit free.
For $s\geq2$, $q_s(K)$ measures the degree of anisotropy of rank~$s$, i.e.,
isotropy results in $q_s(K)=0$, and $q_s(K)=1$ indicates maximal interfacial 
anisotropy.
The anisotropy index $q_2(K)$ can be expressed by the eigenvalues 
$\delta_1(K)$ and $\delta_2(K)$ of the Minkowski tensor $W_1^{0,2}(K)$ in the 
Cartesian representation:
\begin{eqnarray}
  q_2(K)&=& \frac{|\delta_1(K) - \delta_2(K)|}{\delta_1(K) + \delta_2(K)}.
  \label{eq_cmt_q2}
\end{eqnarray}
The fourth-rank index $q_4(K)$ can also be conveniently expressed by the 
components of the Minkowski tensor $W_1^{0,4}(K)$ if the coordinate axes 
are assumed to coincide with the main axes of the system (i.e. the 
eigenvectors of $W_1^{0,2}(K)$):
\begin{eqnarray}
    q_4(K) &=&\frac{|(W_1^{0,4}(K))_{xxxx}+(W_1^{0,4}(K))_{yyyy}-6(W_1^{0,4}(K))_{xxyy}|}{(W_1^{0,4}(K))_{xxxx}+(W_1^{0,4}(K))_{yyyy}+2(W_1^{0,4}(K))_{xxyy}}.
  \label{eq_cmt_q4}  
\end{eqnarray}
We derive these relations by expressing the Cartesian Minkowski tensors 
via IMTs:
\begin{eqnarray}
    W_1^{0,s}(K) = \int_{\mathbb{S}^{1}} u^s \egiK(u) du
    &= \sum_{t=-s}^{s} \egiC_t(K) \int\limits_{0}^{2\pi} u^s(\theta)\exponent{it\theta}d\theta,
  \label{eq_circular_MT_cartesian_from_G}
\end{eqnarray}
where the last equality holds because all terms with $|t|>s$ vanish 
because of the vanishing the Fourier components of $u^s$.
Equation~\eqref{eq_circular_MT_cartesian_from_G}
contains the Fourier coefficients of $u^s(\theta)$, which are 
independent of the domain $K$.
For $s=2$, for example, they are given by
\begin{eqnarray}
  \int\limits_{0}^{2\pi} u^2(\theta)\exponent{is\theta}d\theta=
  \left\{
  \begin{array}{l l}
    \pi \ifiopams\mathds{1} \fi & \mathrm{if\ } s=0,\\
    \frac{\pi}{2} \left({\sigma_3} + i {\sigma_1} \right) & \mathrm{if\ } s=2,\\
    \frac{\pi}{2} \left({\sigma_3} - i {\sigma_1} \right) & \mathrm{if\ } s=-2,\\
    \ifiopams\mathbb{O} \fi & \mathrm{else},
  \end{array}
  \right.
\end{eqnarray}
where \ifiopams$\mathds{1}$ \fi is the unit tensor, \ifiopams$\mathbb{O}$ \fi the zero tensor, and ${\sigma_j}$ are Pauli matrices with
\[
  \sigma_3=\left(
  \begin{array}{c c}
    1 &  0\\
    0 & -1
  \end{array}
  \right)
  \quad \mathrm{and} \quad
  \sigma_1=\left(
  \begin{array}{c c}
    0 &  1\\
    1 &  0
  \end{array}
\right).
\]
Thus, $W_1^{0,2}(K)$ can be expressed by $\egiC_0(K)$, $\egiC_1(K)$, and $\egiC_2(K)$, and vice versa.
From
\begin{eqnarray}
  W_1^{0,2}(K) &=& \egiC_0(K) \pi \mathbf{1} + \egiC_2(K)   \frac{\pi}{2}   \left({\sigma_3} + i {\sigma_1} \right) + \egiC_{-2}(K)   \frac{\pi}{2}   \left({\sigma_3} - i {\sigma_1} \right)\\
  &=& \egiC_0(K)  \pi  \mathbf{1} + \mathrm{Re}[\egiC_2(K)]  {\pi}  {\sigma_3} - \mathrm{Im}[\egiC_2(K)]   {\pi}  {\sigma_1}
\end{eqnarray}
follows that
\begin{eqnarray}
  \egiC_0(K)      &=& \frac{1}{2\pi} \left((W_1^{0,2}(K))_{xx}+(W_1^{0,2}(K))_{yy}\right) = \frac{1}{2\pi}W_1(K),\\
  \mathrm{Re}[\egiC_2(K)] &=& \frac{1}{2\pi} \left((W_1^{0,2}(K))_{xx}-(W_1^{0,2}(K))_{yy}\right),\\
  \mathrm{Im}[\egiC_2(K)] &=& -\frac{1}{\pi} (W_1^{0,2}(K))_{xy}.
  \label{eq_cmt_from_w102}
\end{eqnarray}

\section*{References}

\end{document}